\documentclass{aa}
\usepackage{epsfig}
\usepackage{aalongtable}
\usepackage{txfonts}

\newcommand{\ALi}{$A{\rm (Li)}$\,}
\newcommand{\vsini}{${\rm \it vsini}$\,}

\newcommand{\teff}{T$_{\rm eff}$}

\newcommand{\msun}{$\mathrm{M_{\odot}}$}

\newcommand{\bmv}{B$-$V}

\newcommand{\kms}{\mbox{\rm km\,s$^{-1}$}}

\begin{document}
\title{Rotation and lithium abundance of solar--analog stars
}
   \subtitle{Theoretical analysis of observations}

   \author{J. D. do Nascimento Jr, J.S. da Costa, \and J.R. De Medeiros
}

   \offprints{J.D. do Nascimento Jr., email:dias@dfte.ufrn.br}

\institute{Departamento de F\'{\i}sica Te\'{o}rica e Experimental,
             Universidade Federal do Rio
            Grande do Norte, 59072-970  Natal, R.N., Brazil
                 }
\titlerunning{Lithium and rotation of solar--analog stars}
\date{Received Date: Accepted Date}

\abstract
 \abstract{}{}{}{}{}
\abstract
{Rotational velocity, lithium abundance, and the mass depth of the outer convective zone  are key parameters in the study of the processes at work
in the stellar interior, in particular when examining the poorly understood processes operating in the interior of solar--analog stars.}
{We investigate whether the large dispersion in the observed lithium abundances of solar--analog stars can be explained by the  depth behavior of the 
outer convective zone masses, within the framework of the    standard convection model based on the
local mixing--length theory. We also aims to analyze the link between rotation and lithium abundance in solar--analog stars.}
{We computed a new extensive grid of stellar evolutionary models, applicable to solar--analog stars, for a finely discretized 
set of  mass and  metallicity. From these models, the stellar mass, age, and mass depth of the outer convective zone were estimated for 
117  solar--analog stars, using Teff  and [Fe/H] available in the literature, and the  new HIPPARCOS trigonometric parallax 
measurements.}
{We determine the age and mass of the outer convective zone for a  bona fide sample of 117 solar--analog stars.
No significant one--to--one correlation is found between the computed convection zone mass and published lithium 
abundance, indicating that the large  \ALi dispersion in solar analogs cannot 
be explained by the classical framework of envelope convective mixing 
coupled with lithium depletion at the bottom of the convection zone.}
{These results illustrate the need for an extra--mixing process to explain lithium behavior in solar--analog stars, such as, 
shear mixing caused by differential rotation. To derive a more realistic definition of solar--analog stars, 
as well as solar--twin stars, it seems important  to consider  the inner  physical properties of stars,  such as convection,  hence rotation and magnetic properties.}

\authorrunning{do Nascimento et al.}

\maketitle
\section{Introduction}\label{intro}
The evolutionary behavior of lithium abundance and rotation, and the convective properties of stars depend strongly on stellar mass. 
The genesis of this dependence is of obvious interest in stellar astrophysics, in particular for the study of solar--analog stars,  
but very little is  known about this subject.  For the convective properties, perhaps the foremost difficulty  lies in  the criteria  used to the define 
solar--analog or solar twin,  stars that are  spectroscopically and photometrically identical to the Sun 
(Cayrel de Strobel 1996). Different studies have suggested that the Sun could be (i) an abnormally slow rotator, and (ii) lithium--poor 
by a factor of 10 (Lambert and Reddy 2004) relative to similar solar--type disk stars, which raises the question of whether 
there is anything unusual about the solar rotation rate and Li abundance? King 2005, Takeda et al. 2007, 
Mel\'endez and Ramirez 2007 demostrated that, at least in terms of its lithium content, the Sun is a normal star at its present  evolutionary stage.
\begin{figure*}
\vspace{.2in}
\centerline{\psfig{figure=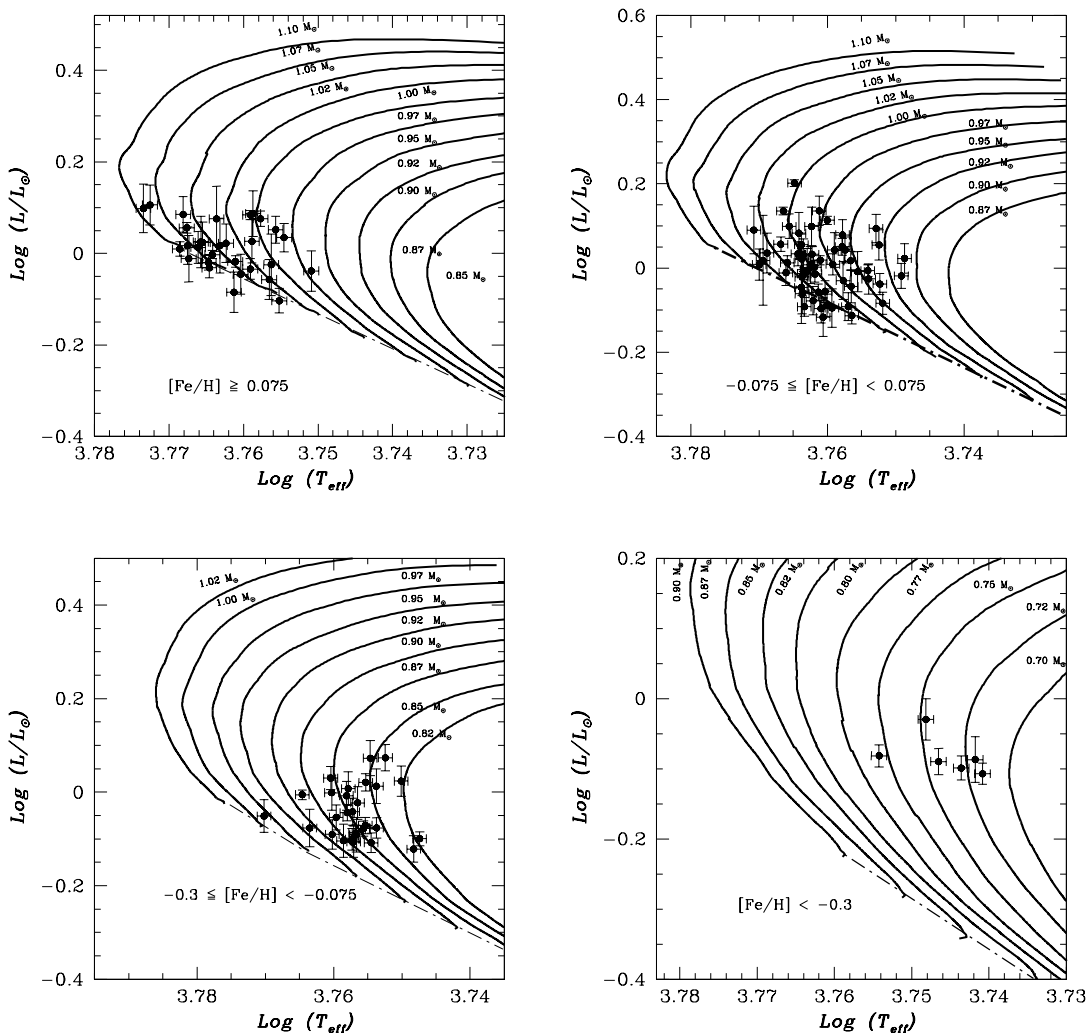, width=6.0truein,height=6.0truein}
\hskip 0.1in}
\caption{The distribution of the analog sample stars in the
Hertzsprung-Russell diagram.  Luminosities and related errors have
been derived from the Hipparcos parallaxes.  The typical error in
\teff~ is  $\pm$ 12~K (Takeda et al.~\cite{tak07}).  Evolutionary
tracks for [Fe/H]=0.15, 0.0, -0.20, and -0.40 and stellar masses
spanning the 0.70 to 1.1 \msun~  mass range.}
\label{hrhip}
\end{figure*}
In spite of dozens of stars having been classified as solar--analogs (e.g: Takeda et al. 2007), only three solar twins are known:  
18 Sco (Porto de Mello \& da Silva 1997), HD 98618 (Mel\'endez et al. 2006), and HIP 100963 (Takeda et al. 2007), all of witch have a Li abundance
higher than the solar value by a factor of beetwen 3 and 6.  For instance, spectroscopic analysis of the solar twin 18 Sco by Porto de Mello and Da
Silva~(\cite{gus97}) demostrated that the atmospheric parameters, chromospheric activity, and UBV colors of this object are 
indistinguishable from  solar values, but that the system has an excess of Sc, V, and heavier elements. In addition, Mel\'endez and Ramirez (2007) 
presented data for two solar twins of low Li 
abundance (HIP 56948 and HIP 73815), pointing to a factor of about 200 in Li depletion relative to that found in meteorites. This finding clearly 
contradicts some standard model predictions (models  without rotation at any depth with the hypothesis that stellar convective 
regions are instantly mixed and that no  chemical transport occurs in the radiative regions) and represents a long-standing puzzle in stellar
astrophysics. Pasquini  et al. (2008) showed that five bona--fide main--sequence stars of the open cluster M 67 are potential 
solar--twins. These stars have low Li content, comparable to the photospheric solar value, probably indicating  similar mixing evolution. In 
addition, these authors confirmed the presence of a large Li spread among the solar--type stars of M~67, showing for the first time that
 extra Li depletion appears only in stars cooler than 6000 K. This behavior indicates that these stars very likely experience a depletion in their Li during 
the main sequence (MS) stage, although their convective zones do not reach sufficiently deep into the stellar interior to meet the zone of Li 
destruction. The scatter in Li abundance in the solar cluster M~67 by a factor of $\sim 10$ around the solar-age (Spite et al. \cite{spi87}; 
Garc\'\i a L\'opez et al. \cite{gar88}) is a solid example  of the disagreement between observations and the theoretical predictions of standard 
models, implying that depletion must be affected by an additional parameter besides mass, age, and chemical composition 
(Pasquini et al.~\cite{pas97}; Jones et al.~\cite{jon99}; Randich et al. 2006).

\begin{figure}
\vspace{.2in}
\centerline{\psfig{figure=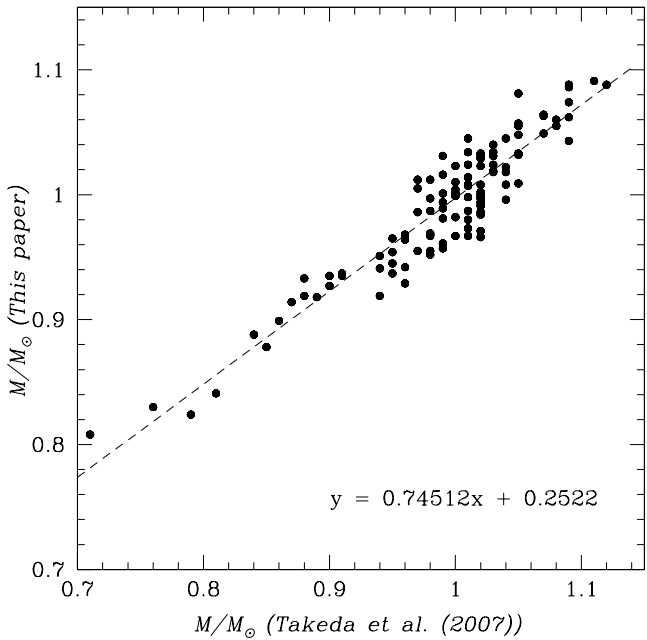, width=3.5truein,height=3.5truein}
\hskip 0.1in}
\caption{Comparison between the  mass determinations of this
study, determined with the Tou\-louse-Geneva
tracks  and those computed by Takeda et al.~(\cite{tak07}).
}
\label{hrcomp}
\end{figure}
\par In the present work, we investigate the influence of mass,~\teff, age, and
mass depth of the outer convective zone on the behavior of Li abundance
and rotational velocity in solar--analog stars and twins. A close examination of some  stellar parameters
(stellar mass and convection--zone mass deepening) may shed new
light on the lithium behavior in this family of stars.

We redetermine the evolutionary status and individual masses for a sample
of 117 solar--analogs, using HIPPARCOS parallaxes and by comparing the observational
Hertzsprung-Russell diagram with evolutionary tracks, following the procedure described in Sect. 2.
As in \cite{donascimento09}, we show that it is possible to precisely determine the 
mass and age of solar analogs and twins using evolutionary models calibrated to reproduce solar 
luminosity, radius, and Li depletion. The characteristics of the working sample are also described in Sect. 2. 
In Sect. 3, the main li\-thium and rotation features are presented, with an analysis of the influence of mass
 and convection zone depth on these parameters. Finally, the main conclusions are outlined in Sect. 4.

\section{Stellar evolutionary models and working sample}

In the next sections, we discuss about the stellar evolutionary models  
ingredients and  the characteristics of the observational data used in this study.
\subsection{Stellar evolutionary models}
\begin{figure}
\vspace{.2in}
\centerline{\psfig{figure=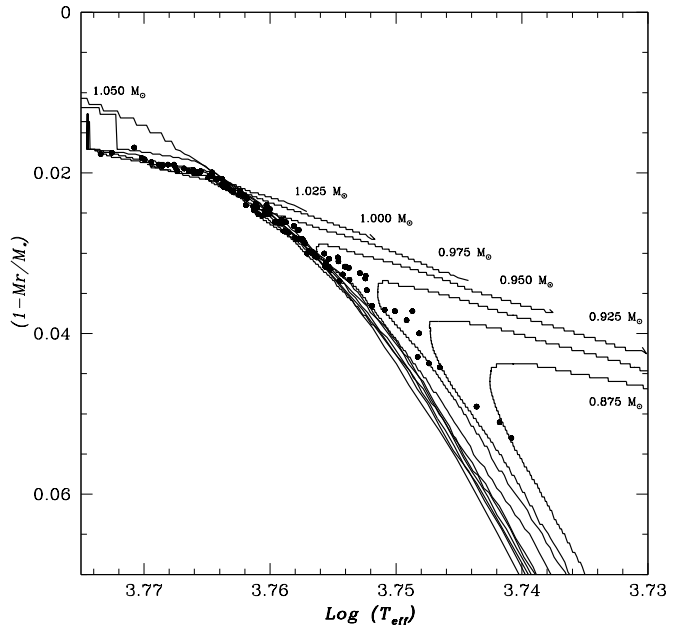, width=3.5truein,height=3.5truein}
\hskip 0.1in}
\caption{Convective zone mass deepening as a function of  decreasing
effective temperature. Models for  $[Fe/H]=0$ and  0.875, 0.900, 0.925,
0.950, 0.975,  1.00, 1.025, 1.050, 1.075,  and 1.100 \msun.  The solid circle represents
the surface convective mass determination of this study as a function of
effective temperature for each star in the sample.}
\label{convteff}
\end{figure}

For the purposes of the present study, evolutionary tracks were computed with the Toulouse--Geneva stellar evolution code TGEC (Hui-Bon-Hoa
\cite{bonhoa07}). Details about the underlying physics of these models can be found in Richard et al. (\cite{oliv96}), do Nascimento et al.~(\cite{donas2000}), 
Hui-Bon-Hoa (\cite{bonhoa07}), and  \cite{donascimento09}.  Here, we provide a short description of the main physical stellar model ingredients.\\

\noindent \textit{Input physics} \vspace{.15in} \\
We used the OPAL2001 equation of state by Rogers and Nayfonov (\cite{rogers02}) and  the radiative opacities by Iglesias \& Rogers
(\cite{igle96}), in addition to the  low temperature atomic and molecular opacities by Alexander \& Ferguson (\cite{alex94}). The nuclear 
reactions are from the analytical formulae of the NACRE (Angulo et al. 1999) compilation, taking into account the three \textit{pp} 
chains and the CNO tricycle, with the \cite{bahcallpinsonneault92} screening routine. Convection is treated according to the 
\cite{bohmvitense58} formalism of the mixing--length theory with $\alpha_p = l/H_p = 1.756515$. For the atmosphere, we used a grey 
atmosphere following the Eddington relation. The abundance variations of the following chemical species were  individually computed in 
the stellar evolution code: H, He, C, N, O, Ne, and Mg. The heavier elements were gathered in Z.
The initial composition followed the Grevesse and Noels (1993) mixture, with initial helium abundance $Y_{ini} = 0.270$.  All models 
included  gravitational settling with diffusion coefficients computed as in \cite{paquette86}. Radiative accelerations were not computed 
here, as we only focus on solar-type stars whose effects are negligible.

Our model grid includes 17 mass tracks spanning the mass range from 0.70 to 1.1 \msun~ for four different metallicities,
($[Fe/H]=$ 0.15, 0.0, -0.20 and -0.40). The evolution was followed from the zero-age main sequence (ZAMS) to the end of hydrogen
exhaustion in the core. Evolution calculations were computed with a short step  to match the effective solar temperature observed and luminosity at the solar age. The model calibration method was based on
Richard et al.~(\cite{oliv96}), as follows: for a 1.00 \msun~star, we calibrated the mixing-length parameter ($\alpha_p$)
and initial helium abundance ($Y_{ini}$) to match the observed solar luminosity and radius at solar age. The observed values that we used were those obtained by Richard et al.~(2004): $L_{\odot}=$ 3.8515 $\pm$  0.0055 $\times$ 10$^{33}$
erg.s$^{-1}$, $R_{\odot}=$ 6.95749 $\pm$ 0.00241 $\times$ 10$^{10}$ cm and
$Age_{\odot}=$
4.57 $\pm$ 0.02 Gyrs. For our best solar model, we obtained L = 3.8541
$\times$ 10$^{33}$ erg.s$^{-1}$ and R = 6.95743 $\times$ 10$^{10}$ cm at age = 4.57 Gyrs.  The input parameters for the other masses were the same as 
those of  the 1.00 \msun~ model. To verify model deviations
in the mass determination, we compared  the evolutionary tracks
computed of this study  with those used by Takeda et al.~(2007)
(evolutionary tracks from Girardi et al. 2000). Both evolutionary tracks
are of solar metallicity, with stellar masses by Takeda et al.~(2007)
covering only three values, namely 0.9,  1.0, and 1.1 \msun.

\subsection{Working sample}

Our analysis is based on the observational data obtained by Takeda et al.~(\cite{tak07}) for a sample of 117 field solar--analog stars selected
from the HIPPARCOS  catalog (ESA 1997), according to the criteria ${\rm V}  <  8.5$, 0.62   $\leq  \bmv  \leq$  0.67 and 4.5  $\leq  M_{\rm V}
 \leq  5.1$. These authors determined lithium abundance from the resonance Li~{\sc i} 6707.8~\AA~doublet, with associated errors around
 0.1~dex, (caused by uncertainties in atmospheric parameters). Rotational velocities were obtained from Nordstr\"om et al. (2004) and Holmberg et al. (2007), with
typical errors of about 1 \kms. The reader is referred to these papers for observational procedures, data reduction, and error analysis. We also 
re--examined the kinematic properties of each star based on the proper motion data taken from the HIPPARCOS catalog and  radial velocity  
measurements, confirming the result found by Takeda et al.~(\cite{tak07}) that all 117 field solar--analog stars selected belong to the normal 
thin-disk population.

\begin{figure*}
\vspace{.2in}
\centerline{\psfig{figure=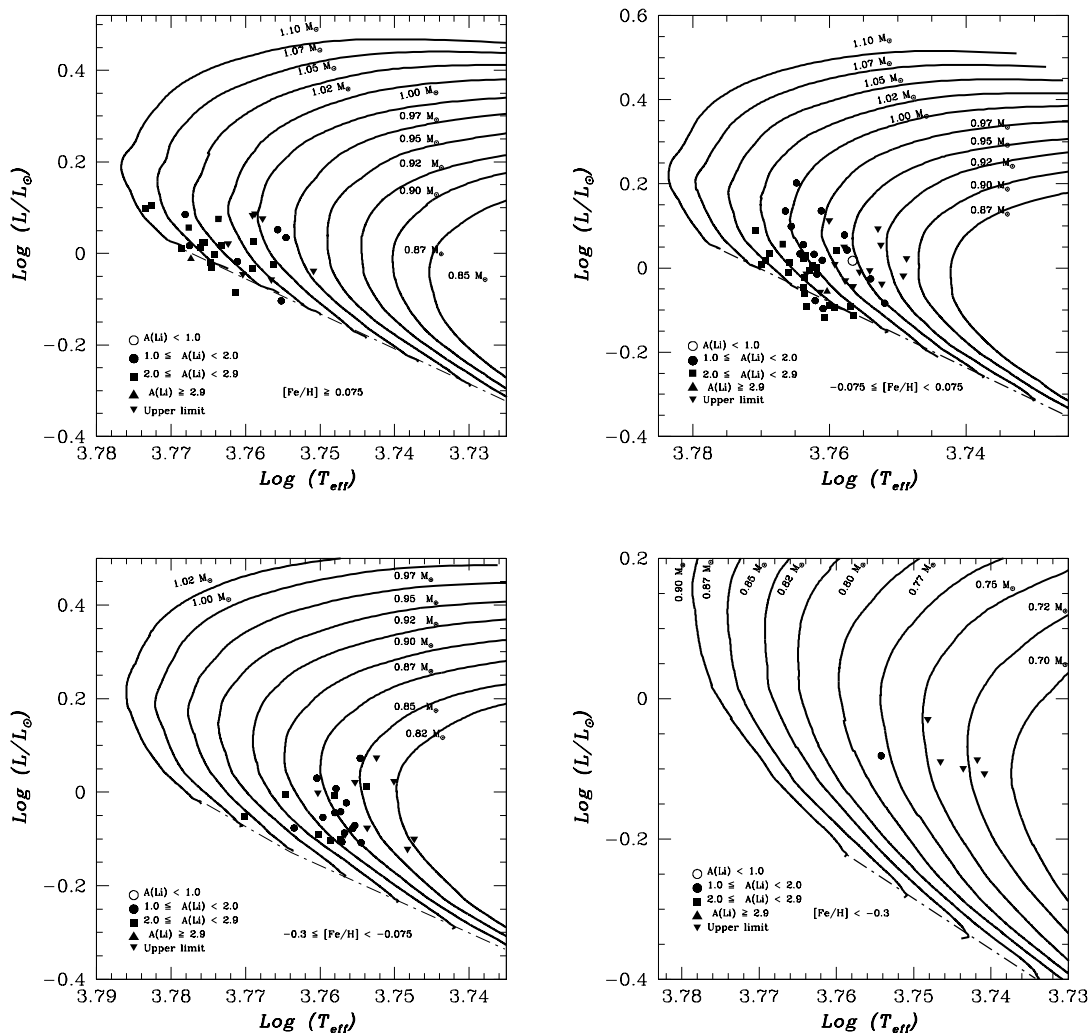,  width=6.0truein,height=6.0truein}
\hskip 0.1in}
\caption{Distribution of  Li abundances in the HR diagram.
The different symbols represent  Li abundances.   Filled inverse triangles represent the upper 
limits Li to the abundances. Evolutionary
tracks, as in Fig.1.  }
\label{hrali}
\end{figure*}
Following the procedure of do Nascimento et al. (\cite{donas2000}), we used the new HIPPARCOS trigonometric parallax measurements to 
precisely locate the objects in the HR diagram. Intrinsic absolute magnitudes M$_{\rm V}$ were derived from the parallaxes, the ${\rm V}$ 
magnitudes also being taken from HIPPARCOS. Stellar luminosity and the associated error were computed from the $\sigma$ error in the parallax.
The uncertainties in luminosity, $\pm0.1$, have an effect of $\pm0.03$ in the determination of the masses. Figure.~\ref{hrhip} shows the HR diagram 
with the evolutionary tracks computed for four different metallicity values ($[Fe/H]$ = 0.15, 0.0, -0.20, and -0.40), which encompasses most of the
stars contained  in the present working sample.

Table~\ref{compa}  shows a comparison between the theoretical luminosities, effective temperatures, and ages of the Toulouse-Geneve evolution code 
(TGEC)  and Girardi et al. (2000) to examine the difference around the observed  effective solar temperature and luminosity. Our  values 
for age, luminosity, and effective temperature are in close agreement with both those of Richard et al. (1996) and helioseismological predictions. 
Some  discrepancies were found between our values and those of  Girardi et al. (2000).

\subsection{Mass, age, and convective mass  determination}

\begin{figure}
\vspace{.2in}
\centerline{\psfig{figure=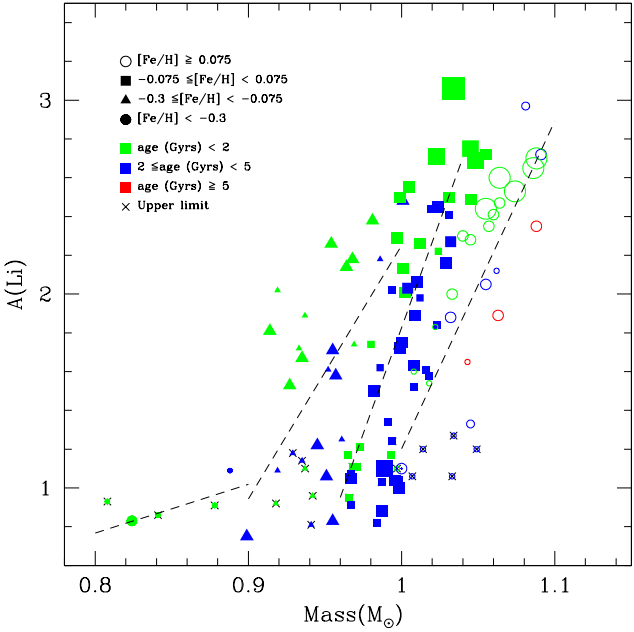, width=3.5truein,height=3.5truein}
\hskip 0.1in}
\caption{Lithium abundance as a  function of  mass (in \msun).
Different symbols stand for the following metallicity intervals: open circle
$\ensuremath {\left [{\rm Fe}/{\rm H}\right ]}\geq 0.075$,
square $-0.075\leq \ensuremath {\left [{\rm Fe}/{\rm H}\right ]} <0.075$,
triangle $-0.3\leq \ensuremath {\left [{\rm Fe}/{\rm H}\right ]} <-0.075$,
solid  circle $\ensuremath {\left [{\rm Fe}/{\rm H}\right ]} <-0.3$.
 Upper  limit Li abundances are represented by x.
The Sun is also displayed  for comparative purposes. The symbol size indicate
the projected rotational velocities (\vsini). }
\label{alimass}
\end{figure}
       We computed a careful star-by-star model for each star  in our sample,  with  metallicity that matching that observed for 
each solar analog.  The errors in \teff, luminosity, and [Fe/H] correspond to a mean error of  0.05 \msun~ in the
mass. As a consistency check, a comparison between our mass determination and those obtained by Takeda et al.~(\cite{tak07}) is shown in Fig. 2. 
Except for some minor discrepancies for stars with masses lower than 0.8 \msun~bacause of a lack of models used by Takeda et al.~(\cite{tak07}), there is 
generally close agreement between the  two sets of masses. Stellar ages were inferred simultaneously with the masses, from the evolutionary tracks. 
The error in these ages, due to the uncertainties in \teff, luminosity, and [Fe/H], is about 15\% and depends strongly on  parallax 
uncertainties. Stars located close to the ZAMS have much larger errors. A comparison with age values computed by Holmberg  et al. (2007) and
Takeda et al.~(\cite{tak07}), again finds no significant differences. Figure ~\ref{convteff} shows the convective zone mass deepening as a function of the decreasing effective temperature for 0.875, 0.900, 0.925, 0.950, 0.975,
1.00, 1.025, 1.050, 1.075, and 1.100 \msun, where only the  solar metallicity models are considered. In this figure, solid circles represent the convection 
mass zone, ${(1-Mr/M_{\rm *}})$, for the stars of the present sample of solar--analogs,  determined in the scope of this study, where we have taken into
consideration the metallicity of each star. Solar--analog stars typically, exhibit outer convective zone masses of around 0.02 \msun~(see ${1-Mr/M_{\rm *}}$ 
values in Table 2).  A close look at Fig.~\ref{convteff} also indicates that the onset of convection zone deepening (in mass)  is a strong function of 
stellar mass at the end of the pre-main-sequence phase. The depth of  the outer convective zone  in main-sequence stars depends primarily on mass.
 Deepening of the outer convective zone is also sensitive to metallicity. Quantitatively, a reduction in the initial  metallicity $[Fe/H]_0$  from 0.0  
 to -0.4  should lead to a decrease in the outer convective zone  ${(1-Mr/M_{\rm *}})$  from  0.022 to 0.004 at the Sun's age. Lithium data can be used to
 constrain the mass and metallicity-dependent process of  the outer convective zone.

\begin{figure}
\vspace{.2in}
\centerline{\psfig{figure=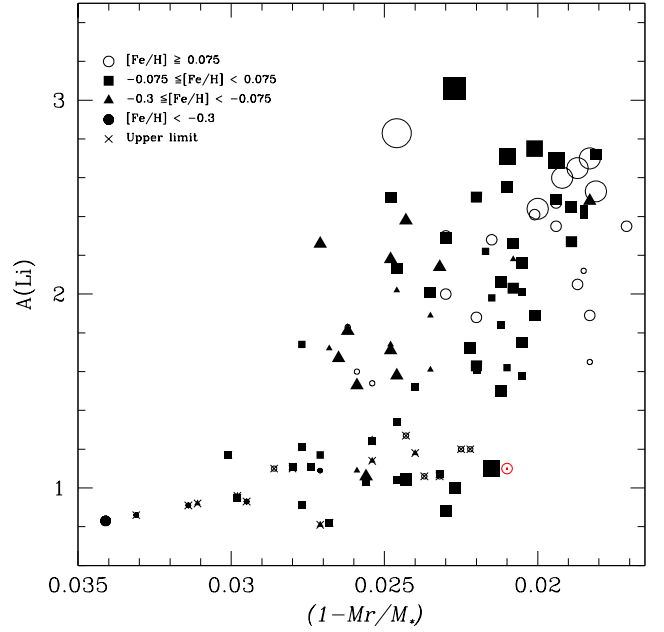, width=3.5truein,height=3.5truein}
\hskip 0.1in}
\caption{Lithium abundance as a function of convection zone mass
deepening ${(1-Mr/M_{\rm *}})$. Stars are grouped into metallicity intervals,
as in Fig 5. Upper  limit Li abundances are represented by x.  The Sun is also represented for comparative purposes.The symbol size indicate
the projected rotational velocities (\vsini). }
\label{alimzc}
\end{figure}

Lithium depletion  in solar-twins caused by non-standard  mixing is strongly mass--dependent (do Nascimento et al. 2009).  A detailed comparison of the mass  
determinations of Takeda et al.~(\cite{tak07}) and Nordstr\"om et al. (2004) reveals some discrepancies,  such as   HIP 45325.   The masses 
obtained by  Takeda et al.~(\cite{tak07}) are systematically higher than the values obtained by  Nordstr\"om et al. (2004). These discrepancies may be 
related to the completeness of the set  of evolutionary models used by both authors.  We recalculated as precisely as possible the mass of  all 
sample stars. In this study, we used a set of homogeneous evolutionary tracks for main--sequence stars, computed for several masses and metal 
abundances, as discussed in Sect. 2.2. Mass determination was evaluated taking into account the metallicity effects,  using the [Fe/H] 
values published by Takeda et al.~(\cite{tak07}).

\section{Results and discussion}

We now discuss the relashionsip between the lithium abundance of solar--analog stars and their position in the HR diagram, verifying in particular the influence  
of the mass and  outer convective zone depth on the letter. We consider in addition the rotational behavior of these stars. In 
addition, we show the important role of inner stellar properties in defining solar--analog stars and twins.

\subsection{Main lithium features} \label{analysis}

  Figure ~\ref{hrali} shows the distribution of Li abundance for
solar--analog stars in the HR diagram, with stars grouped into four different metallicity intervals, namely [Fe/H] $\geq$ 0.075,  -0.075
$\leq$  [Fe/H]  $<$ 0.075, -0.3  $\leq$   [Fe/H]  $<$ -0.075, and [Fe/H] $<$ -0.3. The evolutionary tracks shown
are those computed in the scope of the present work. A number of  features can be observed in this figure:
 {\bf (i)} Stars with masses $<$ 0.82~M$_{\odot}$ tend to have the lowest lithium content, most likely indicating that the 
 same  depletion level  occurred in both the pre-main sequence and the early main sequence. 
 {\bf (ii)} Stars with masses between 0.82 and 1.1~M$_{\odot}$ have a large
scatter in their lithium content, possibly reflecting the different levels of lithium depletion.
This star-to-star scatter in Li abundance was also observed in some solar-like field stars (Soderblom  et al. 1993;  
Jones et al.~\cite{jon99}) and in members of the solar--age cluster M~67 (Pasquini  et al. 2008).

The lithium scatter observed for solar--analogs indicates, in part, that these stars have also had different convection 
histories, which  depend strongly on stellar mass, \teff~, metallicity ,and age. The classical interpretation 
is that Li has been destroyed in the stellar interior  by proton-proton reactions 
at the bottom of the outer convection, i.e., at $T_{bcz} > 2.5 \times 10^{6} K$.  Since the depth of the convection zone
depends primarily on stellar mass, a correlation between Li abundance and mass is then expected for a given
metallicity. This is  observed in   Figs.~\ref{alimass} and ~\ref{alimzc}, which show the distribution of lithium abundance as a function of stellar mass and 
convection zone mass deepening ${(1-Mr/M_{\rm *}})$, respectively. In these figures, stars are grouped into the following  metallicity
 ranges: open circles $\ensuremath {\left [{\rm Fe}/{\rm H}\right ]}\geq 0.075$, squares $-0.075\leq \ensuremath {\left [{\rm Fe}/{\rm H}\right ]} <0.075$,
triangles $-0.3\leq \ensuremath {\left [{\rm Fe}/{\rm H}\right ]} <-0.075$, solid circles $\ensuremath {\left [{\rm Fe}/{\rm H}\right ]} <-0.3$.
The Sun, with \ALi~ = 1.1 (Grevesse and Sauval 1998) and ${(1-Mr/M_{\rm *}})$ = 0.021 (see Table 2), is also displayed in both figures for 
comparative purposes.

\begin{figure}
\vspace{.2in}
\centerline{\psfig{figure=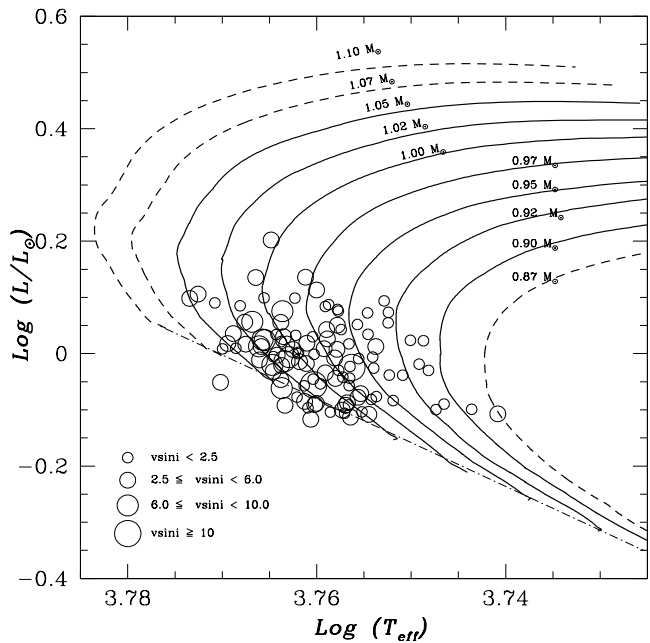, width=3.5truein,height=3.5truein}
\hskip 0.1in}
\caption{Distribution of projected rotational velocity measurements
({\it vsini}, in \kms) in the  HR diagram. The different symbols represent
the rotational velocities. The evolutionary tracks are similar to those as in Fig.1.}
\label{hrvsini}
\end{figure}

 Figures  \ref{alimass} and \ref{alimzc}  show two interesting properties as solar--analog stars. Although the masses of the present sample of
solar--analogs appears to span a narrow range of values, namely  masses between 0.82 and 1.1~M$_{\odot}$, one observes a strong dispersion in 
the distribution of lithium abundance versus stellar mass, once finely discretized sets of mass are considered. For masses $M  \geq 0.85~M_{\odot}$, 
in particular, a wide range of about 2.5 orders of  magnitude of \ALi is observed in Fig~\ref{alimass}, a dispersion also observed at any given mass. 

Since convection zone deepening depends primarily on both stellar mass and age, the \ALi pattern in  
 Fig.~\ref{alimass} should again parallel that observed in Fig.~\ref{alimzc}, at least in relation to the \ALi dispersion. A 
comparison between Figs.~\ref{alimass} and ~\ref{alimzc} indicates that stars with the smallest masses have the lowest Li abundances. This feature reinforces the claim that the root--cause 
of  the low lithium content observed in these stars is primarily related to their previous evolutionary 
history and that other parameters in addition to stellar mass and metallicity affect the degree of depletion.

\subsection{Lithium and rotation relationship in solar--analog stars}

The behavior of the rotational velocity is of obvious importance to our understanding of the lithium-rotation relationships in solar--analog stars 
because it is largely accepted that rotation has a major influence on Li abundances. Different studies, again suggest that the Sun may be an abnormally slow rotator (Lambert and Reddy 2004), compared to similar solar--type disk stars. 
Unfortunately, there are two major difficulties in this analysis caused by stellar rotational velocity. First, the true rotation of the 
vast majority of solar--analogs is unknown, because the main detection procedure currently in use gives provides the the projected 
rotational velocity, namely the minimum stellar rotation (\vsini). The definitive  rotation rate can only be derived  using a photometry procedure. 
To date, the literature lists the rotational period for only a small percentage of the identified solar-analog stars. For instance, Messina and 
Guinan (2004) provide the photometric rotation period for 6 apparent solar--analogues,  all the stars with periods between 2.6 and 9.21 days 
rotating more rapidly than the Sun. Unfortunately, only one of these 6 solar-analog stars is listed in the literature. In addition,  Butler et al. (1998)
measure a period of about 30 d for the solar analog HD 187123. In spite of these difficulties, the preliminary study by 
Takeda et al. (2007) illustrates that solar analogues with lower rotational velocity (\vsini) also have lower Li abundances. This conclusion is based 
on measurements of line widths, from which theses authors estimated the \vsini+macroturbulence of each star. This result is confirmed by 
Gonzalez (2008), using  \vsini  measurements,  separated from the macroturbulence, for a sample of stars with planets.

We now revisit  the rotational behavior analyses of solar--analog stars by considering the same sample of 
solar--analogues studied by Takeda et al. (2007), but now using  \vsini  from Nordstr\"om et al. (2004),  which corresponds to the difference in
rotation from the macroturbulence.  Figure ~\ref{hrvsini} shows the distribution of the projected rotational velocity ({\it vsini}) in 
the  HR diagram, for the aforementioned   stellar sample, with   evolutionary tracks computed in the scope of the present 
work. We can clearly observe,  an important scatter in the distribution of {\it vsini} with 
mass, in particularly among stars of mass  $\geq$ 0.95 \msun . For more massive stars, one also observes 
significant scatter in {\it vsini} at a given mass. This latter result implies that parameters others than stellar mass and 
\teff~ may also be controlling the rotation of solar--analog stars. 

The strong dispersion in \ALi at a given stellar mass or convection zone depth, for $(1-Mr/M_{\rm *}) \leq 0.03$ and 
mass $> 0.9$ \msun, as shown in the previous section, indicates that  Li depletion depends on parameters other than stellar  mass,
such as age and early rotational history. To analyze in detail the role of rotation and age in  this context, we illustrate in Fig.~\ref{alimass}
the distribution of \ALi versus mass, with the stars segregated  by metallicity, age ,and \vsini.  First, it is clear that solar--analogs
show the same trends between lithium content and rotation observed for other stellar families,  where the largest \vsini is associated with the largest
\ALi. For instance, except for one star, HD 166435  (HIP 88945), all the stars with enhanced rotation exhibit a large \ALi. In contrast, among 
slow rotators one observes a large spread in lithium content, from the smallest \ALi $\sim -0.8$ to the largest \ALi $\sim 3$ values. Furthermore,
this figure shows an inhabited region for low mass stars, in the sense that for masses lower than about 0.84 \msun, stars with larger \ALi than
about 1 and enhanced rotation seem unusual, a pattern that is clearly associated with stellar age.

\section{Conclusions} \label{sect_results}

     In the past decade, different studies of low-mass stars have provided important information about the physical properties of solar--analog
stars.  Spectroscopic data have offered important border conditions to choose good candidates, based on the surface composition of
photometrically solar--analog stars.  Nevertheless, the important scatter in the distribution of different physical parameters, such as lithium
abundance and rotation, indicates, in the broad sense, that information regarding surface convection and angular momentum may be 
essential to the classification of solar--analog and solar twins stars. To investigate the evolutionary status, mass, and any correlation
between Li abundance, convection, and rotation of solar--analog G dwarf stars, we computed a grid of hundreds of evolutionary models 
for stars of different metallicity and mass range, from 0.7 to 1.1~\msun.  These  evolutionary tracks were computed using the Tou\-louse-Geneva 
code with  updated physical inputs, as  described in do Nascimento et al. (2009).  Our analysis of  lithium abundance in a sample of 
solar--analog  stars found different degrees of lithium depletion, a large scatter in the abundance being observed for stars over  a narrow range 
of mass and metallicity. Low Li abundance among solar--analog stars strongly supports the hypothesis that these stars  have depleted Li 
during the MS phases. These results illustrate the need for an extra--mixing process to explain lithium behavior in solar--analog stars, such as, 
shear mixing caused by differential rotation proposed by Bouvier (2008).

The aforementioned  dispersion in the lithium abundance of solar--analog stars at a given age and mass may also reflect their different rotational histories.
In spite of the short range of mass in the  solar--analog stars, this study  shows how sensitive \ALi scatter  is as a function of stellar mass,
metallicity, age and, \teff. It cannot however be excluded that the Galactic cosmic Li abundance dispersion could contributes to the scattering in 
lithium abundance.  Even if solar--analog stars were chosen according  to  their spectroscopic and photometric similarity with the Sun, our 
conclusion in this study reinforces  the need for information about the stellar interior. Furthermore, the Li depletion observed in the solar--analog
stars  and the large spread in Li abundances among the five known solar twins cannot be explained only by standard convective mixing.
At a given \teff~  or age, a small percentage  difference in stellar mass, and  consequently in  the depth (in mass) of the stellar 
convective envelope, can produce a  \ALi scatter, which tends to increase with stellar mass and \teff. This illustrates that in producing a more realistic 
definition of solar--analog and solar--twin stars, it seems important to consider the inner physical properties of stars,  such as the depth of the convective envelope mass and consequently rotation and magnetic properties. 
Finally, we note that  asteroseismology offers a unique opportunity to study the extension of outer convection zones of
solar--analogs. This study and a determination of the chemical abundance pattern  of solar--analog and solar--twins stars could
 play an important role in answering the fundamental question about how normal the Sun as a star really is.
\begin{acknowledgements}

This research  made use of the SIMBAD data base, operated at CDS, Strasbourg, France. JDN and JRM are 
research fellows of the   CNPq  Brazilian agency. Research activities of the Stellar Board at the 
Federal University of Rio Grande do Norte are supported by continuous grants from CNPq and 
FAPERN Brazilian Agencies. We thank the anonymous referee for the useful
comments and suggestions.

\end{acknowledgements}

{}

\begin{table}
\caption{Comparison between  effective temperature and luminosity
at the solar age   from the Toulouse-Geneve evolution code TGEC   and
Girardi et al. (2000).
}\label{compa}
\begin{tabular}{ccccc}\\ \hline
Model          & log L      &  log(\teff) & age  &    \\ \hline
TGEC           & -0.00033  &  3.76212   & $ 4.5767739 \times 10^9$  &    \\
TGEC           & 0.00020  &  3.76215   & $ 4.5917307 \times 10^9$  &    \\
Girardi et al. & -0.021     &  3.760     & $4.02403\times 10^9$     & \\
Girardi et al. & 0.023      &  3.762     & $5.20367\times 10^9$     & \\ \hline
\end{tabular}
\end{table}

\begin{table*}
\small
\caption{Parameters of the working sample. Derived masses,  outer convective zone ${(1-Mr/M_{\rm *}})$, temperature at the bottom of the convective zone 
T$_{bcz}$, age, and  luminosities for our program stars. $^{a}$Grevesse and Sauval (1998) $^{b}$, Valenti and Fischer (2005) 
}
\label{Tabbase}																							
\begin{flushleft}																							
\begin{tabular}{cclrccrrrrl}																							
\noalign{\smallskip}																							
\hline\\[-2mm]																							
HIP	&	HD	&	log(\teff)&$\log(L/L_\odot)$	&	M$/$\msun	&	${(1-Mr/M_{\rm	*}})$	&	T$_{bcz}$	(10$^{6}$	K)	&	age	(Gyrs)	&\ALi	&	[Fe/H]	&{\it	vsini}\\[2mm]		
\hline\\[-2mm]																							
\hline\\[-2mm]																							
1499	&	1461	&	3.758	&	0.076$\pm$0.017	&	1.03	&	0.023	&	2.25	&	3.34	&	$<$ 1.06	&	0.197	&	2	\\	
1598	&	1562	&	3.755	&	-0.071$\pm$0.017	&	0.91	&	0.026	&	2.2	&	1.26	&		1.81	&	-0.272	&	3	\\	
1803	&	1835	&	3.765	&	-0.020$\pm$0.016	&	1.09	&	0.019	&	2.19	&	0.94	&		2.65	&	0.239	&	7	\\	
4290	&	5294	&	3.757	&	-0.101$\pm$0.023	&	0.97	&	0.025	&	2.23	&	0.93	&		2.18	&	-0.116	&	3	\\	
5176	&	6512	&	3.768	&	0.017$\pm$0.041	&	1.06	&	0.018	&	1.98	&	5.76	&		1.89	&	0.185	&	3	\\	
6405	&	8262	&	3.758	&	-0.044$\pm$0.018	&	0.96	&	0.025	&	2.21	&	2	&		1.71	&	-0.137	&	4	\\	
6455	&	8406	&	3.757	&	-0.106$\pm$0.033	&	0.97	&	0.025	&	2.22	&	1.81	&		1.74	&	-0.088	&	2	\\	
7244	&	9472	&	3.76	&	-0.088$\pm$0.027	&	1.0	     &	0.023	&	2.21	&	1.24	&		2.29	&	-0.039	&	4	\\	
7585	&	9986	&	3.762	&	0.033$\pm$0.018	&	1.02	&	0.021	&	2.17	&	2.84	&		1.84	&	0.074	&	1	\\	
7902	&	10145	&	3.749	&	-0.019$\pm$0.030	&	0.97	&	0.03	&	2.36	&	1.33	&	$<$	0.95	&	-0.008	&	1	\\	
7918	&	10307	&	3.766	&	0.135$\pm$0.009	&	1.01	&	0.02	&	2.08	&	4.64	&		1.89	&	0.012	&	3	\\	
8486	&	11131	&	3.764	&	-0.046$\pm$0.085	&	1.01	&	0.021	&	2.16	&	1.02	&		2.55	&	-0.061	&	4	\\	
9172	&	11926	&	3.761	&	-0.117$\pm$0.045	&	1.03	&	0.022	&	2.22	&	0.86	&		2.5	&	0.063	&	5	\\	
9349	&	12264	&	3.762	&	0.007$\pm$0.044	&	1.01	&	0.021	&	2.16	&	2.52	&		2.06	&	0.01	&	3	\\	
9519	&	12484	&	3.767	&	-0.011$\pm$0.051	&	1.08	&	0.018	&	2.14	&	0.42	&		2.97	&	0.139	&	-	\\	
9829	&	12846	&	3.747	&	-0.090$\pm$0.019	&	0.88	&	0.031	&	2.27	&	1.09	&	$<$	0.91	&	-0.306	&	2	\\	
10321	&	13507	&	3.756	&	-0.113$\pm$0.020	&	1.0 	&	0.025	&	2.27	&	0.59	&		2.5	&	-0.005	&	3	\\	
11728	&	15632	&	3.757	&	-0.044$\pm$0.039	&	0.99	&	0.026	&	2.27	&	2.8	&	$<$	1.03	&	0.023	&	1	\\	
12067	&	15851	&	3.757	&	-0.057$\pm$0.043	&	1.03	&	0.024	&	2.3	&	2.6	&	$<$	1.27	&	0.202	&	1	\\	
14614	&	19518	&	3.758	&	0.008$\pm$0.036	&	0.96	&	0.025	&	2.2	&	2.58	&		1.58	&	-0.122	&	3	\\	
14623	&	19632	&	3.759	&	-0.034$\pm$0.023	&	1.03	&	0.023	&	2.26	&	1.61	&		2	&	0.122	&	4	\\	
15062	&	20065	&	3.759	&	-0.104$\pm$0.036	&	0.92	&	0.025	&	2.16	&	1.35	&		2.02	&	-0.287	&	2	\\	
15442	&	20619	&	3.754	&	-0.108$\pm$0.021	&	0.94	&	0.027	&	2.23	&	1.37	&		1.67	&	-0.188	&	3	\\	
16405	&	21774	&	3.759	&	0.087$\pm$0.050	&	1.05	&	0.022	&	2.23	&	3.47	&	$<$	1.2	&	0.263	&	1	\\	
17336	&	23052	&	3.754	&	-0.076$\pm$0.022	&	0.94	&	0.027	&	2.24	&	2.58	&	$<$	0.81	&	-0.126	&	1	\\	
18261	&	24552	&	3.769	&	0.036$\pm$0.047	&	1.03	&	0.019	&	2.09	&	3.06	&		2.27	&	0.024	&	3	\\	
19793	&	26736	&	3.765	&	0.024$\pm$0.044	&	1.07	&	0.018	&	2.13	&	1.63	&		2.53	&	0.187	&	6	\\	
19911	&	26990	&	3.754	&	0.013$\pm$0.037	&	0.95	&	0.027	&	2.27	&	0.65	&		2.26	&	-0.127	&	5	\\	
19925	&	27063	&	3.761	&	-0.096$\pm$0.034	&	1.02	&	0.022	&	2.18	&	3.26	&		1.61	&	0.071	&	2	\\	
20441	&	27685	&	3.761	&	-0.085$\pm$0.044	&	1.05	&	0.022	&	2.22	&	1.44	&		2.28	&	0.13	&	3	\\	
20719	&	28068	&	3.766	&	0.025$\pm$0.056	&	1.06	&	0.019	&	2.17	&	1.21	&		2.6	&	0.134	&	6	\\	
20741	&	28099	&	3.763	&	0.017$\pm$0.060	&	1.06	&	0.02	&	2.19	&	1.38	&		2.41	&	0.163	&	4	\\	
20752	&	28192	&	3.773	&	0.106$\pm$0.044	&	1.09	&	0.013	&	1.9	&	2.02	&		2.72	&	0.155	&	5	\\	
21165	&	27757	&	3.76	&	0.030$\pm$0.025	&	0.95	&	0.024	&	2.16	&	2.65	&		1.61	&	-0.16	&	2	\\	
21172	&	28821	&	3.75	&	0.024$\pm$0.032	&	0.94	&	0.03	&	2.33	&	1.37	&	$<$	0.96	&	-0.103	&	1	\\	
22203	&	30246	&	3.759	&	0.026$\pm$0.047	&	1.04	&	0.023	&	2.27	&	0.95	&		2.3	&	0.133	&	3	\\	
23530	&	31864	&	3.748	&	-0.122$\pm$0.028	&	0.9	&	0.031	&	2.29	&	1.28	&	$<$	0.75	&	-0.238	&	-	\\	
25002	&	35041	&	3.758	&	-0.008$\pm$0.030	&	0.98	&	0.024	&	2.23	&	0.78	&		2.38	&	-0.083	&	5	\\	
25414	&	35073	&	3.751	&	-0.038$\pm$0.044	&	1.0 	&	0.029	&	2.37	&	1.44	&	$<$	1.1	&	0.097	&	1	\\	
25670	&	36152	&	3.76	&	-0.045$\pm$0.043	&	1.01	&	0.023	&	2.18	&	3.98	&	$<$	1.2	&	0.099	&	1	\\	
26381	&	37124	&	3.742	&	-0.087$\pm$0.033	&	0.83	&	0.034	&	2.26	&	1.01	&	$<$	0.66	&	-0.447	&	-	\\	
27435	&	38858	&	3.756	&	-0.078$\pm$0.016	&	0.93	&	0.026	&	2.2	&	1.91	&		1.53	&	-0.215	&	3	\\	
29432	&	42618	&	3.757	&	-0.087$\pm$0.017	&	0.95	&	0.026	&	2.22	&	2.79	&		1.06	&	-0.117	&	4	\\	
31965	&	47309	&	3.761	&	0.136$\pm$0.035	&	1.0	&	0.023	&	2.16	&	4.64	&		1	&	0.048	&	3	\\	
32673	&	49178	&	3.758	&	-0.030$\pm$0.045	&	1.0	&	0.025	&	2.24	&	3.41	&	$<$	1.04	&	0.06	&	2	\\	
33932	&	49985	&	3.77	&	-0.051$\pm$0.035	&	1.0	&	0.018	&	2.04	&	2.56	&		2.48	&	-0.118	&	4	\\	
35185	&	56202	&	3.763	&	-0.006$\pm$0.052	&	1.02	&	0.021	&	2.18	&	0.67	&		2.71	&	-0.003	&	7	\\	
35265	&	56124	&	3.764	&	0.022$\pm$0.025	&	1.0	&	0.021	&	2.12	&	2.82	&		2.01	&	-0.018	&	1	\\	
36512	&	59711	&	3.757	&	-0.042$\pm$0.029	&	0.96	&	0.025	&	2.23	&	2.62	&		1.25	&	-0.091	&	2	\\	
38647	&	64324	&	3.757	&	-0.092$\pm$0.033	&	1.0	&	0.025	&	2.26	&	1.01	&		2.13	&	0.01	&	3	\\	
38747	&	64942	&	3.764	&	-0.061$\pm$0.047	&	1.05	&	0.02	&	2.18	&	0.66	&		2.75	&	0.065	&	6	\\	
38853	&	65080	&	3.771	&	0.090$\pm$0.056	&	1.02	&	0.019	&	2.06	&	2.55	&		2.44	&	-0.052	&	1	\\	
39506	&	66573	&	3.748	&	-0.030$\pm$0.029	&	0.81	&	0.03	&	2.13	&	1.28	&	$<$	0.93	&	-0.62	&	2	\\	
39822	&	66171	&	3.76	&	-0.001$\pm$0.037	&	0.93	&	0.024	&	2.15	&	3.28	&	$<$	1.18	&	-0.218	&	2	\\	
40118	&	68017	&	3.744	&	-0.099$\pm$0.017	&	0.84	&	0.033	&	2.26	&	0.98	&	$<$	0.86	&	-0.42	&	2	\\	
40133	&	68168	&	3.756	&	0.052$\pm$0.030	&	1.02	&	0.025	&	2.31	&	1.73	&		1.54	&	0.122	&	2	\\	
41184	&	70516	&	3.756	&	-0.025$\pm$0.067	&	1.03	&	0.025	&	2.31	&	0.29	&		2.83	&	0.106	&	13	\\	
41526	&	71227	&	3.764	&	0.030$\pm$0.039	&	1.0 	&	0.021	&	2.13	&	2.64	&		2.03	&	-0.017	&	3	\\	
42333	&	73350	&	3.765	&	-0.032$\pm$0.021	&	1.06	&	0.019	&	2.16	&	1.94	&		2.35	&	0.144	&	5	\\
42575	&	73393	&	3.754	&	-0.026$\pm$0.036	&	0.99	&	0.027	&	2.32	&	1.91	&		1.17	&	0.058	&	1	\\	
43297	&	75302	&	3.755	&	-0.104$\pm$0.027	&	1.01	&	0.026	&	2.31	&	1.56	&		1.6	&	0.083	&	2	\\	
43557	&	75767	&	3.764	&	0.056$\pm$0.025	&	0.98	&	0.021	&	2.11	&	4.18	&		1.5	&	-0.06	&	4	\\	
43726	&	76151	&	3.761	&	-0.018$\pm$0.013	&	1.03	&	0.022	&	2.21	&	2.32	&		1.88	&	0.114	&	3	\\	
44324	&	77006	&	3.77	&	0.008$\pm$0.037	&	1.03	&	0.019	&	2.07	&	2.76	&		2.41	&	-0.01	&	1	\\	
44997	&	78660	&	3.756	&	-0.009$\pm$0.048	&	0.99	&	0.025	&	2.27	&	2.57	&	$<$1.24	&	0.044	&	2	\\	
45325	&	79282	&	3.773	&	0.098$\pm$0.053	&	1.09	&	0.017	&	1.95	&	5.32	&		2.35	&	0.178	&	3	\\																								
\hline																							
\end{tabular}																							
\\																							
\end{flushleft}																							
\end{table*}

\begin{table*}																							
(continued)																							
\begin{flushleft}																							
\begin{tabular}{cclrccrrrrl}																							
\noalign{\smallskip}																							
\hline\\[-2mm]																							
HIP	&	HD	&	log(\teff)&$\log(L/L_\odot)$	&	M$/$\msun	&	${(1-Mr/M_{\rm	*}})$	&	T$_{bcz}$	(10$^{6}$	K)	&	age	(Gyrs)	&\ALi	&	[Fe/H]	&{\it	vsini}\\[2mm]		
\hline\\[-2mm]																							
\hline\\[-2mm]	

46903	&	82460	&	3.759	&	-0.095$\pm$0.046	&	0.99	&	0.023	&	2.21	&	1.88	&		2.02	&	-0.025	&	-	\\	
49580	&	87680	&	3.762	&	-0.077$\pm$0.034	&	1.01	&	0.022	&	2.17	&	2.44	&		1.98	&	0.023	&	1	\\	
49586	&	87666	&	3.762	&	0.022$\pm$0.042	&	1.05	&	0.02	&	2.09	&	4.77	&	$<$	1.33	&	0.2	&	-	\\	
49728	&	88084	&	3.759	&	0.009$\pm$0.025	&	0.97	&	0.024	&	2.19	&	3.43	&	$<$	1.05	&	-0.068	&	3	\\	
49756	&	88072	&	3.757	&	0.043$\pm$0.031	&	0.99	&	0.025	&	2.24	&	2.88	&		1.34	&	0.019	&	2	\\	
50505	&	89269	&	3.747	&	-0.100$\pm$0.015	&	0.92	&	0.031	&	2.32	&	1.11	&	$<$	0.92	&	-0.169	&	1	\\	
51178	&	90494	&	3.763	&	-0.077$\pm$0.040	&	0.95	&	0.023	&	2.11	&	3.89	&	$<$	1.22	&	-0.174	&	-	\\	
53721	&	95128	&	3.765	&	0.202$\pm$0.008	&	1.0 	&	0.021	&	2.11	&	3.68	&		1.75	&	-0.019	&	3	\\	
54375	&	96497	&	3.764	&	0.075$\pm$0.071	&	1.06	&	0.02	&	2.19	&	1.44	&		2.44	&	0.136	&	6	\\	
55459	&	98618	&	3.764	&	0.034$\pm$0.028	&	1.02	&	0.021	&	2.12	&	3.81	&		1.58	&	0.066	&	1	\\	
55868	&	99505	&	3.76	&	-0.091$\pm$0.031	&	0.96	&	0.023	&	2.17	&	1.44	&		2.14	&	-0.148	&	4	\\	
56948	&	101364	&	3.762	&	0.099$\pm$0.046	&	-	&	-	&	-	&	-	&		-	&	0.02	&	1	\\	
59589	&	106210	&	3.752	&	-0.038$\pm$0.025	&	0.97	&	0.028	&	2.31	&	1.66	&	$<$	1.21	&	-0.013	&	2	\\	
59610	&	106252	&	3.766	&	0.099$\pm$0.029	&	0.99	&	0.021	&	2.09	&	4.39	&		1.62	&	-0.062	&	1	\\	
62175	&	110869	&	3.755	&	0.034$\pm$0.031	&	1.02	&	0.026	&	2.35	&	1.02	&		1.83	&	0.133	&	1	\\	
62816	&	111938	&	3.764	&	-0.021$\pm$0.044	&	1.03	&	0.021	&	2.16	&	2.26	&		2.16	&	0.062	&	4	\\	
63048	&	112257	&	3.752	&	0.055$\pm$0.036	&	0.97	&	0.028	&	2.32	&	1.78	&	$<$	1.11	&	-0.016	&	2	\\	
63636	&	113319	&	3.763	&	-0.092$\pm$0.023	&	1.01	&	0.021	&	2.15	&	1.88	&		2.26	&	-0.01	&	4	\\	
64150	&	114174	&	3.758	&	0.051$\pm$0.020	&	1.0 	&	0.024	&	2.22	&	3.75	&	$<$	1.04	&	0.054	&	3	\\
64747	&	115349	&	3.757	&	-0.091$\pm$0.037	&	0.94	&	0.025	&	2.19	&	2.76	&	$<$	1.14	&	-0.178	&	2	\\	
70319	&	126053	&	3.754	&	-0.081$\pm$0.016	&	0.89	&	0.027	&	2.18	&	2.27	&		1.09	&	-0.327	&	2	\\	
72604	&	131042	&	3.752	&	0.073$\pm$0.028	&	0.94	&	0.028	&	2.27	&	1.9	&	$<$	1.1	&	-0.142	&	1	\\	
73815	&	133600	&	3.764	&	0.083$\pm$0.049	&	-	&	-	&	-	&	-	&		-	&	0.01	&	-	\\	
75676	&	138004	&	3.761	&	-0.057$\pm$0.020	&	0.97	&	0.023	&	2.16	&	4.13	&	$<$	1.07	&	-0.075	&	2	\\	
76114	&	138573	&	3.757	&	0.017$\pm$0.023	&	0.97	&	0.028	&	2.3	&	2.34	&		0.91	&	-0.023	&	2	\\	
77749	&	142072	&	3.766	&	0.013$\pm$0.035	&	1.09	&	0.018	&	2.17	&	0.97	&		2.7	&	0.223	&	7	\\	
78217	&	144061	&	3.76	&	-0.054$\pm$0.027	&	0.94	&	0.024	&	2.15	&	1.99	&		1.89	&	-0.224	&	1	\\	
79672	&	146233	&	3.761	&	0.018$\pm$0.011	&	1.01	&	0.022	&	2.17	&	3.3	&		1.63	&	0.039	&	3	\\	
85042	&	157347	&	3.754	&	-0.006$\pm$0.014	&	0.98	&	0.027	&	2.29	&	2.62	&	$<$	0.82	&	0.034	&	1	\\	
85810	&	159222	&	3.768	&	0.056$\pm$0.011	&	1.06	&	0.019	&	2.09	&	3.89	&		2.05	&	0.149	&	3	\\	
88194	&	164595	&	3.755	&	0.021$\pm$0.018	&	0.96	&	0.026	&	2.24	&	2.99	&	$<$	0.83	&	-0.083	&	-	\\	
88945	&	166435	&	3.763	&	-0.013$\pm$0.015	&	0.99	&	0.022	&	2.11	&	4.59	&	$<$	1.1	&	-0.01	&	8	\\	
89282	&	167389	&	3.766	&	0.013$\pm$0.017	&	1.02	&	0.019	&	2.1	&	2.09	&		2.45	&	-0.002	&	3	\\	
89474	&	168009	&	3.76	&	0.113$\pm$0.010	&	0.99	&	0.023	&	2.16	&	4.41	&	$<$	0.88	&	0.006	&	3	\\	
89912	&	168874	&	3.767	&	0.057$\pm$0.017	&	1.05	&	0.019	&	2.15	&	0.93	&		2.69	&	0.041	&	7	\\	
90004	&	168746	&	3.749	&	0.023$\pm$0.035	&	0.97	&	0.03	&	2.37	&	1.03	&	$<$	1.17	&	-0.015	&	2	\\	
91287	&	171665	&	3.752	&	-0.084$\pm$0.026	&	0.98	&	0.028	&	2.32	&	0.92	&		1.74	&	-0.007	&	2	\\	
96184	&	184403	&	3.768	&	0.085$\pm$0.039	&	1.04	&	0.018	&	1.92	&	6.7	&		1.65	&	0.132	&	1	\\	
96395	&	185414	&	3.765	&	-0.005$\pm$0.010	&	0.99	&	0.021	&	2.12	&	2.32	&		2.18	&	-0.1	&	1	\\	
96402	&	184768	&	3.753	&	0.094$\pm$0.034	&	0.97	&	0.027	&	2.29	&	1.96	&	$<$	1.11	&	-0.033	&	1	\\	
96901	&	186427	&	3.759	&	0.084$\pm$0.010	&	1.01	&	0.024	&	2.22	&	3.29	&	$<$	1.06	&	0.081	&	2	\\	
96948	&	186104	&	3.758	&	0.079$\pm$0.037	&	1.01	&	0.024	&	2.24	&	2.52	&		1.52	&	0.07	&	2	\\	
97420	&	187237	&	3.762	&	0.002$\pm$0.016	&	1.02	&	0.022	&	2.2	&	1.74	&		2.22	&	0.053	&	2	\\	
98921	&	190771	&	3.764	&	-0.003$\pm$0.009	&	1.06	&	0.019	&	2.17	&	1.54	&		2.47	&	0.167	&	4	\\	
100963	&	195034	&	3.762	&	-0.014$\pm$0.019	&	1.0	&	0.022	&	2.17	&	2.89	&		1.72	&	-0.002	&	3	\\	
104075	&	200746	&	3.769	&	0.018$\pm$0.107	&	1.06	&	0.018	&	2.11	&	1.24	&		2.72	&	0.051	&	5	\\	
109110	&	209779	&	3.766	&	-0.011$\pm$0.031	&	1.05	&	0.019	&	2.14	&	1.61	&		2.49	&	0.071	&	5	\\	
110205	&	211786	&	3.756	&	-0.023$\pm$0.035	&	0.92	&	0.026	&	2.19	&	2.68	&		1.09	&	-0.229	&	1	\\	
112504	&	215696	&	3.759	&	0.043$\pm$0.041	&	1.0 	&	0.024	&	2.22	&	1.63	&		2.01	&	0.007	&	3	\\	
113579	&	217343	&	3.76	&	-0.056$\pm$0.025	&	1.03	&	0.023	&	2.24	&	0.15	&		3.06	&	0.054	&	13	\\	
113989	&	218209	&	3.741	&	-0.107$\pm$0.015	&	0.82	&	0.034	&	2.25	&	0.9	&	$<$	0.83	&	-0.462	&	3	\\	
115715	&	220821	&	3.755	&	0.072$\pm$0.038	&	0.93	&	0.027	&	2.24	&	1.32	&		1.72	&	-0.192	&	2	\\	
116613	&	222143	&	3.769	&	0.010$\pm$0.016	&	1.06	&	0.019	&	2.07	&	4.16	&		2.12	&	0.156	&	2	\\	
Sun	&	--	&	3.762$^{a}$	&	0	&	1.0	&	0.021	&	--	&	--	&		1.1$^{a}$	&	--	&	1.7$^{b}$	\\	
[2mm]																							
\hline																							
\end{tabular}																							
\\																							
\end{flushleft}																							
\end{table*}

\end{document}